\documentclass[11pt,a4paper]{article}
\usepackage{epsf}

\def\figuresize{12cm}

\title{Hartree-Fock based diagonalization: an efficient method for simulating disordered
 interacting electrons}

\author{Thomas Vojta, Frank Epperlein and Michael Schreiber\\
          \small Institut f\"ur Physik, Technische Universit\"at, D-09107 Chemnitz, Germany}

\date{\small version August 31, compiled \today}

\begin{document}
\maketitle

\begin{abstract}
We present an efficient numerical method for simulating the low-energy properties of 
disordered many-particle systems. The method which is based on the quantum-chemical 
configuration interaction approach consists in diagonalizing the Hamiltonian in an energetically
truncated basis build of  the low-energy states of the corresponding Hartree-Fock Hamiltonian. 
As an example we investigate the quantum Coulomb glass, a model
of spinless electrons in a random potential interacting via long-range Coulomb
interaction. We find that the Coulomb interaction increases the conductance of strongly disordered
systems but reduces the conductance of  weakly disordered systems.
\end{abstract}

The numerical simulation of disordered many-particle systems is one of the most complicated 
problems in computational condensed matter physics. First, the size of the 
Hilbert space to be considered grows exponentially with the system size. Second, the presence
of disorder requires the simulation of many samples with different disorder configurations 
in order to
obtain averages or distribution functions of physical quantities. In the case of disordered interacting
electrons the problem is made worse by the long-range character of the Coulomb interaction
which has to be retained, at least for a correct description of  the insulating phase.

The simulation methods applied to disordered many-particle systems can be roughly 
divided into two classes. On the one hand, methods like Hartree-Fock \cite{hf,epper_hf} have 
been used to reduce the system to an effective single-particle system. This overcomes
the problem of the Hilbert space growing exponentially with system size.
The resulting methods permit the simulation of rather large systems ($>10^3$ sites)
but the approximations involved are uncontrolled and can usually not be 
improved systematically. 

On the other hand, there are several methods which give numerically exact results 
or which can be taken, at least in principle, to arbitrary accuracy.
However, most of these methods are severely restricted when simulating
disordered interacting electrons. 
Exact diagonalization \cite{exact,epper_exact} works only for very small systems 
(with up to about $4 \times 4$ lattice sites). 
For one-dimensional systems the density-matrix renormalization 
group method \cite{dmrg} is a very efficient tool to obtain the low-energy properties.
It is, however, less effective in higher dimensions; and it is also not capable of handling
the long-range Coulomb interaction which is important in the insulating phase.
Quantum Monte-Carlo \cite{qmc} methods are another means of simulating disordered many-particle
systems. They are very effective for Bosons at finite temperatures. Very low temperatures are,
however, hard to reach. Moreover, simulations of Fermions suffer from the notorious sign problem
(although this turned out to be less severe in the presence of disorder).

In this paper we present an alternative method for simulating disordered interacting
electrons which we call the Hartree-Fock based diagonalization (HFD). 
It is based on the quantum chemical configuration interaction \cite{CI}
approach adapted for disordered lattice models. The main idea is to diagonalize
the Hamiltonian in a subspace of the Hilbert space spanned by 
the low-energy eigenstates of the Hartree-Fock approximation of the 
Hamiltonian.
The HFD method consists of 3 steps: 
(i) solve the Hartree-Fock approximation of the Hamiltonian which is still a non-trivial
   {\em disordered} single-particle problem,
(ii) use a Monte-Carlo algorithm to find the low-energy many-particle Hartree-Fock states, and
(iii) diagonalize the Hamiltonian in the basis formed by these states and calculate the 
   observables. 
The efficiency of the HFD method is due to the fact that the Hartree-Fock 
states are 
comparatively close in character to the exact eigenstates in the entire
parameter space. Thus it works well for all parameters while related methods
based on non-interacting or classical eigenstates \cite{efros95,talamantes96}
instead of Hartree-Fock states are restricted to small parameter regions.

In the following we will illustrate the application of the HFD method on the example of the 
quantum Coulomb glass, a model of interacting spinless electrons in a random 
potential. It is defined on a regular hypercubic lattice with $M=L^d$ ($d$ is the spatial dimensionality) 
sites occupied by $N=K M$ electrons ($0\!<\!K\!<\!1$). To ensure charge neutrality
each lattice site carries a compensating positive charge of  $Ke$. The Hamiltonian
is given by
\begin{equation}
H =  -t  \sum_{\langle ij\rangle} (c_i^\dagger c_j + c_j^\dagger c_i) +
       \sum_i \varphi_i  n_i + \frac{1}{2}\sum_{i\not=j}(n_i-K)(n_j-K)U_{ij}
\label{eq:Hamiltonian}
\end{equation}
where $c_i^\dagger$ and $c_i$ are the electron creation and annihilation operators
at site $i$, respectively,  and $\langle ij \rangle$ denotes all pairs of nearest 
neighbor sites.
$t$ denotes the strength of the hopping term and $n_i$ is the occupation number of site $i$.
$U_{ij} = e^2/r_{ij}$ denotes the Coulomb interaction which we parametrize by its value $U$
between nearest neighbor sites. For a correct description of the insulating phase the Coulomb 
interaction has to be kept long-ranged, since screening breaks down in the insulator. 
The random potential values $\varphi_i$ are chosen 
independently from a box distribution of width $2 W_0$ and zero mean.
For $U_{ij}=0$ the quantum Coulomb glass becomes identical to the Anderson model of
localization  and for $t=0$ it turns into  the classical Coulomb glass.

We now turn to a more  detailed description of the HFD method for the quantum 
Coulomb glass. For each disorder configuration the first step consists of numerically 
diagonalizing the Hartree-Fock approximation
\begin{eqnarray}
H_{\rm HF} = &-t  & \sum_{\langle ij\rangle}  (c_i^\dagger c_j + c_j^\dagger c_i) 
+  \sum_i  \varphi_i  n_i \nonumber \\ 
 &+ & \sum_{i\not=j}  n_i ~ U_{ij} \langle n_j -K \rangle  
- \sum_{i,j} c_i^\dagger c_j ~ U_{ij} \langle c_j^\dagger c_i \rangle,
\label{eq:HF}
\end{eqnarray}
of the Hamiltonian as described in Ref. \cite{epper_hf}. Here
$\langle \ldots \rangle$ represents the expectation value with respect to
the Hartree-Fock ground state which has to be determined self-consistently. 
This calculation results in an orthonormal set of single-particle 
Hartree-Fock states
$|\psi_\nu\rangle=b_\nu^\dagger|0\rangle=\sum_i S_{\nu i}c_i^\dagger|0\rangle$ 

In the second step of the method we construct many-particle states,  i.e. Slater determinants, 
\begin{equation}
  |\{\nu\} \rangle = b_{\nu_1}^\dagger \ldots b_{\nu_N}^\dagger | 0 \rangle~.
\label{eq:MP_states}
\end{equation}
Note that for the two limiting cases mentioned above, 
i.e for the Anderson model of localization and 
for the classical Coulomb glass, these states are also eigenstates of the full
Hamiltonian (\ref{eq:Hamiltonian}).
We then determine which of the many-particle states 
(\ref{eq:MP_states}) have the lowest expectation values 
$\langle \{\nu\}|H_{HF}| \{\nu\}\rangle$ of the energy. Since the total number
of states is too high for a complete enumeration we employ a Monte-Carlo method.
It is based on the thermal cycling method \cite{cycling} in which the system is 
repeatedly heated and cooled. In addition, at the end of each cycle a systematic 
local search around the current configuration is performed.
The low-energy many-particle states  found in this way
span the sub-space of the Hilbert space relevant for the low-energy
properties. Its dimension $B$ determines the accuracy of the results.

The third step consists of transforming the Hamiltonian from the original site-representation
to the Hartree-Fock-representation by means of the unitary transformation 
$b_\nu^\dagger=\sum_i S_{\nu i}c_i^\dagger$ and calculating the matrix elements
$\langle \{\nu\}|H| \{\mu\}\rangle$. The resulting Hamiltonian matrix $H_{\{\nu\}\{\mu\}}$
of dimension $B\times B$ is then diagonalized using standard library routines.
Note that $H_{\{\nu\}\{\mu\}}$ is usually {\em not} very sparse: if  $| \{\nu\}\rangle$ and
$| \{\mu\}\rangle$ differ in the occupation of at most 4 single-particle states, the
matrix element  is non-zero. Moreover, number and position of the non-zero 
matrix elements differ between different disorder configurations. Thus, specialized
codes for sparse matrices will not increase the performance significantly.
In order to investigate physical observables we transform their operators to
the Hartree-Fock representation. This is usually faster than transforming the states
back to site or momentum representation.

\begin{figure}
  \epsfxsize=\figuresize
  \centerline{\epsffile{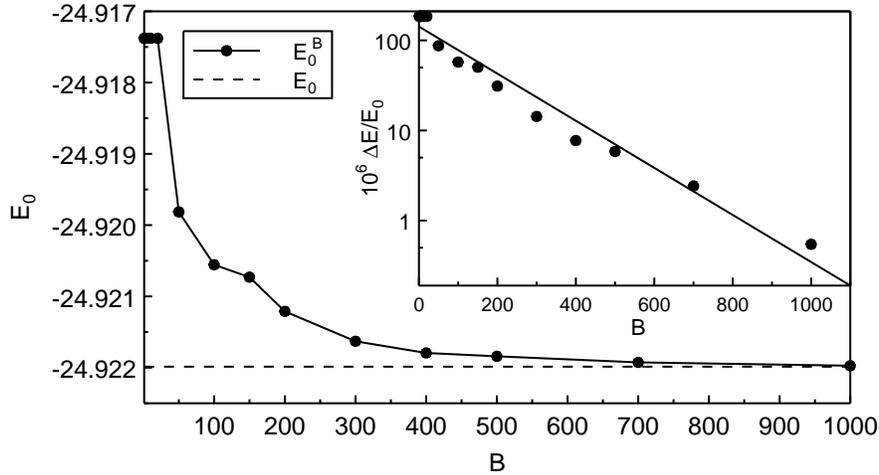}}
  \caption{Dependence of the ground state energy $E_0^B$ and its 
              error $(E_0^B-E_0)/E_0$ 
              on the size $B$ of the basis used for a system of 8 electrons 
              on 16 sites, $W=1, t=0.1, U=1$. The solid line in the inset
              is a fit to an
              exponential law.}
  \label{fig:ground_state_energy}
\end{figure}

\begin{figure}
  \epsfxsize=\figuresize
  \centerline{\epsffile{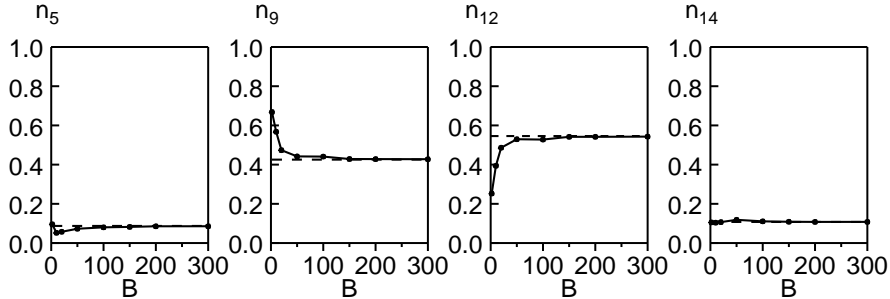}}
  \caption{Occupation numbers $\langle n_i \rangle$ vs. basis size $B$
               for the system of Fig. \ref{fig:ground_state_energy}.}
  \label{fig:occ}
\end{figure}

In order to test the method and to check the dependence of the results on 
the size $B$ of the basis we
carried out extensive simulations for systems with $4 \times 4$ sites and compared the results
to those of exact diagonalizations which are not too time-consuming for spinless electrons 
at this size.
We first investigated the dependence of  the ground state energy $E_0^B$ on $B$ and compared it
to the exact result $E_0$. A typical result is presented in Fig. \ref{fig:ground_state_energy}.
As usual the ground state energy is not very sensitive to the accuracy of the 
approximation. Already the relative energy error of Hartree-Fock is as low as
$10^{-4}$.
Keeping a basis size of 300 within the HFD method reduces the error by
a factor of 10. Further increasing the basis size to 1000 (which is still
less than 10\% of the total Hilbert space) gives a relative  error of less than $10^{-6}$.
Since knowing the energy is not sufficient to judge the quality of the ground state we also
studied the overlap between approximate and exact ground state as well as several 
ground state expectation values. Fig.\ \ref{fig:occ} shows the
convergence of the occupation numbers as a function of basis size $B$
for the same system as in Fig.\ \ref{fig:ground_state_energy}.
While some of the occupation numbers have significant errors within Hartree-Fock
approximation, the HFD method with a basis size of 100 gives all occupation numbers
with a satisfactory accuracy of better than $10^{-2}$.

After having established the method we now show calculations of the transport 
properties. We consider 
the question whether the electron-electron interactions
lead to an enhancement or to a reduction of  the conductance in a system of 
disordered electrons. This question has reattracted a lot of attention after
experiments revealed indications of an unexpected metal-insulator transition in
two dimensions \cite{2DMIT}. 

The conductance is calculated from the Kubo-Greenwood 
formula \cite{kubo_greenwood} which relates it to the
current-current correlation function in the ground state. Using the spectral
representation of the correlation function the real (dissipative) part 
of the conductance (in units of the quantum conductance $e^2/h$)
is obtained as 
\begin{equation}
 \Re ~ G^{xx}(\omega) = \frac {2 \pi^2}   { \omega}  L^{d-2} \sum_{\alpha} |\langle 0 | j^x|\alpha \rangle |^2 
     \delta(\omega+E_0-E_{\alpha})
\label{eq:kubo}
\end{equation}
where $j^x$ is the $x$ component of the current operator and $\alpha$ denotes the eigenstates
of the Hamiltonian.  The finite life time $\tau$ of the eigenstates in a real d.c.\ transport experiment
results in an inhomogeneous broadening $\gamma = 1/\tau$
of the $\delta$ functions in the Kubo-Greenwood formula. Here we have
chosen $\gamma=0.05$ which is of the order of the single-particle level spacing.
According to eq.\ (\ref{eq:kubo}) the accuracy of the conductance depends not only on
the accuracy of the ground state but also on those of the excited states. We therefore 
also carried out convergence tests on the conductance itself, following the lines discussed
above. 

\begin{figure}[ht]
  \epsfxsize=\figuresize
  \centerline{\epsffile{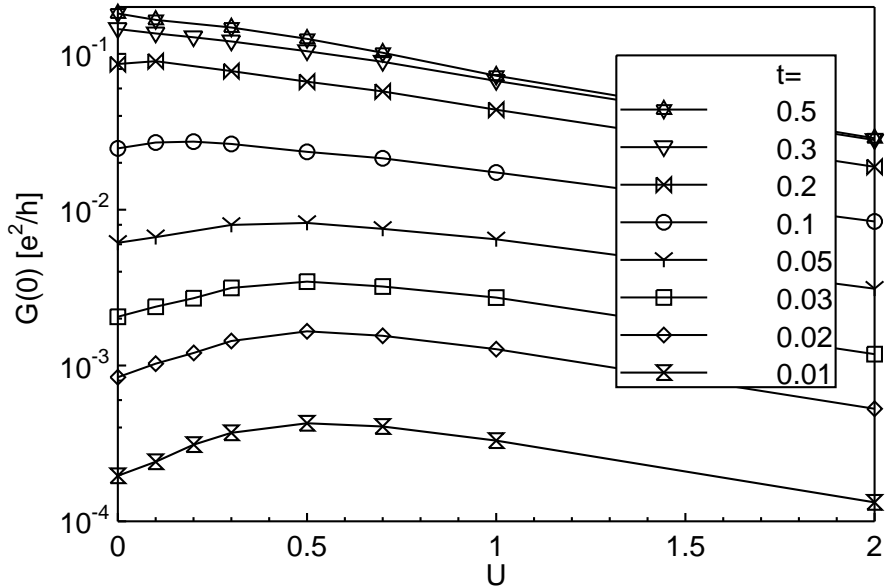}}
  \caption{Logarithmically averaged (400 samples) d.c. conductance for systems of 
              $5 \times 5$ lattice sites and 12 electrons for different $U$ and $t$. The disorder
              strength is fixed to 1.   The basis size within the HFD method was $B=500$.}
  \label{fig:cond}
\end{figure}

We now discuss the main results for the d.c.\ conductance of two-dim\-en\-sional systems.
In Fig.\ \ref{fig:cond} we show the extrapolated d.c. conductances for systems
of $5 \times 5$ lattice sites containing 12 electrons for different values of kinetic energy and
interaction strength. 
The data show that weak electron-electron interactions enhance the conductance for the case of very
small kinetic energy (i.e. large disorder). In this regime the dominant effect of the interactions is that
electron-electron scattering destroys the phase coherence responsible for Anderson localization.
In contrast, for higher kinetic energy or sufficiently strong interactions the dominant effects of the 
interactions are a reduction of the charge fluctuations and an increase in the effective random
potential both of which lead to a reduction of the conductance. Analogous simulations with
different system size and filling factor show the same qualitative dependence of the conductance
on the interaction strength.

To summarize, we have presented an effective method, the Hartree-Fock based diagonalization
(HFD),
for the numerical simulation of disordered many-particle systems. As an example we have applied 
it to the calculation
of transport properties in the quantum Coulomb glass model of interacting electrons in a random
potential. Further results obtained by the 
HFD method on the transport properties of the quantum Coulomb glass 
in one, two and three dimensions can be found in Refs. \cite{letter,giessen,jerusalem}.

This work was supported in part by the German Research Foundation.

\end{document}